\documentclass[showpacs,preprintnumbers,amsmath,amssymb,3p]{elsarticle}
\usepackage{graphicx}
\usepackage{rotating}
\usepackage{csquotes}
\usepackage{bm}
\usepackage{amssymb}
\usepackage{amsmath}
\usepackage{natbib}
\usepackage{dcolumn,multicol}
\makeatletter

\newenvironment{figurehere}
  {\def\@captype{figure}}
  {}
\makeatother
\begin{document}
\begin{frontmatter}

\title{Stochastic modeling for trajectories drift in the ocean: Application of Density Clustering Algorithm}

\author{E. Shchekinova$^{1,2}$, Y. Kumkar$^{2}$}

\address{$^1$ Southern Federal University, Institute of Physics,
Stachki, 194, 344110,Rostov on Don, Russia}
\ead{elena.shchekinova@gmail.com}
\address{$^2$ Euro-Mediterranean Center for Climate Change,
via Augusto Imperatore 16, 73100 Lecce, Italy}

\date{\today}
\begin{abstract}
The aim of this study is to address the effects of wind--induced drift on a floating sea objects using high--resolution ocean forecast data and atmospheric data. Two applications of stochastic Leeway model for prediction of trajectories drift in the Mediterranean sea are presented: long--term simulation of sea drifters in the western Adriatic sea ($21.06.2009-23.06.2009$) and numerical reconstruction of the Elba accident ($21.06.2009-23.06.2009$). Long--term simulations in the western Adriatic sea are performed using wind data from the European Center for Medium--Range Weather Forecast (ECMWF) and currents from the Adriatic Forecasting System (AFS). An algorithm of spatial clustering is proposed to identify the most probable search areas with a high density of drifters. The results are compared for different simulation scenarios using different categories of drifters and forcing fields. The reconstruction of sea object drift near to the Elba Island is performed using surface currents from the Mediterranean Forecasting System (MFS) and atmospheric forcing fields from the ECMWF. The results showed that draft--limited to an upper surface drifters more closely reproduced target trajectory during the accident.
\end{abstract}


\end{frontmatter}
\begin{multicols}{2}
\section{Introduction}\label{sec.1}
The task of developing fast and reliable computer--assisted methods for search and rescue (SAR) operations becomes increasingly important due to increased sea traffic and higher risk of emergencies at sea \cite{FrostStone2001}. 

In previous decades the existing SAR systems relied mostly on climatological means as well as were lacking high resolution environmental data  \cite{FrostStone2001,AllenPlourde1999}. Advances in a high resolution operational forecast in the Mediterranean sea \cite{Davidson2009,Pinardi2003,Oddo2006,Tonani2008} provided a framework for a more accurate prediction of drifters forecast in this area. Also the inclusion of assimilation schemes in ocean forecast models \cite{Dobricic2004} helped to improve a prediction of time--dependent currents and wind with respect to earlier models. 

A large databank of windage characteristics of sea objects was collected over the years \cite{AllenPlourde1999, Allen2005, Allen2010} that uncovered a linear relationship between drift velocity and applied surface wind for different categories of sea drifters. The field observations were used in the context of trajectory prediction using the approach of stochastic Leeway drift \cite{Breivik2008,Breivik2012}. The Leeway model \cite{Breivik2008} produced final positions of drifters based on their windage characteristics \cite{AllenPlourde1999, Allen2005}. Additionally the modeling approach enabled prediction of a search area in terms of a probability density. However, the model does not classify the drifters according to their temporal and spatial behavior. While backtracking simulations could provide a full classification of spatially separated trajectories time limitation could be an issue in the SAR operations and a faster and simpler computational alternative should be provided.     

Here we proposed a computational approach for characterization of drifters based on a density clustering algorithm \cite{Ester1996}. It was designed to improve qualitative information about the stochastic outcome of the Leeway simulations and to increase the prediction probability. The spatial clustering localizes the areas of higher probability of drifters and finds possible outliers. Also the procedure sorts out ensemble of drifters according to their temporal behavior.    

We used the modeling framework based on the explicit parameterization of an object drift by wind \cite{Breivik2008,Daniel2002}. Two applications of the stochastic Leeway model \cite{Breivik2008,Breivik2012} for tracking of small size surface drifters were studied: ($1$) long--term simulation of sea drifters in the western Adriatic sea ($21.06.2009-23.06.2009$) and ($2$) numerical reconstruction of the Elba accident ($21.06.2009-23.06.2009$). In the first study experiments were performed with drifters released in the vicinity of the known hyperbolic structure of the flow near to the Gargano peninsula in the western Adriatic sea. The spatial sorting of drifters was studied by the determination of distinct clusters. Different categories of drifters and geophysical forcing were tested as well. In the second study qualitative comparison between the behavior of reconstructed trajectories of drifters and known information about the Elba accident was performed using ensembles of stochastic simulations. In both applications the clustering method provided improved probability estimate of final search areas via the sorting of drifters according to their spatial distribution.  
\section{Model}\label{sec.2}

\subsection{Model description}
The Leeway drift is defined as a drift of floating object with respect to ambient current under the influence of wind and waves \cite{Breivik2012,Chapline1960,Anderson1998}.
According to the definition the leeway velocity is determined from a difference between drift velocity $\mathbf{V_{dr}}$ and Eulerian current velocity $\mathbf{V_E}$ of a sea object:
\begin{equation}
\mathbf{L}=\mathbf{V_{dr}}-\mathbf{V_E},\label{leewaydrift}
\end{equation}
where $\mathbf{L}$ is the Leeway velocity. 
The integration trajectory is obtained by calculating the drifters path according to formula:
\begin{equation}
\mathbf{dr}=\mathbf{V_{dr}} dt+\mathbf{d\epsilon} =(\mathbf{L}+\mathbf{V_E})dt+\mathbf{d\epsilon} , \label{eqmotion3}
\end{equation}
where $dt$ is the time step, $\mathbf{d\epsilon}$ represents a diffusive component of velocity due to sub--scale variability of ocean currents that is below the given model resolution, $\mathbf{V_E}$ is the Eulerian current taken from the ocean model. The Leeway drift $\mathbf{L}$ is parametrized by $10$~m atmospheric wind $W_{10}$ (see Appendix Eq.~\ref{Leewaycoeff}). Since in field campaigns the contribution of the Stokes drift to the motion of a drifter was difficult to separate from the effect of wind \cite{AllenPlourde1999}. In the present modeling paradigm a sea dominated by weak waves is considered and the waves are assumed to be alighed with the wind. The influence of Stokes drift is implicitly included through Leeway coefficients (see Appendix Eq.~\ref{Leewaycoeff}). The model accounts for the action of wind on small drifters and neglect damping and excitation by waves \cite{Breivik2008,Sorgard1998}.

Since in general shape and size of a sea object is not regular it moves at a certain angle to the wind direction \cite{Anderson1998}. Therefore, the Leeway velocity is decomposed into the downwind (DWL) and crosswind (CWL) Leeway components which are projections of Leeway drift on wind direction $\mathbf{W_{10}}$ \cite{Breivik2008,Anderson1998}. Also due to uncertainty of initial orientation of a drifter with respect to wind an initial ensemble is generated with a half of drifters oriented left and remaining half right with respect to wind direction \cite{Breivik2008}. Because of the existing variability of wind, currents and some perturbation of the drifter motion its orientation can abruptly change. The change of orientation is introduced via the alternation of the sign of the CWL component. In this study probability of change of sign equal to $4 \%$ per integration step \cite{Allen2005}.  

Since a finite size of geographical domain was considered the drifters could reach the boundary of the domain or approach the coastal shoreline. In the former case they were marked as off--grid and removed from further consideration. Stranded drifters were identified according to a high--resolution coastline contour \cite{Wessel1996}.  

Often during emergency situations at sea exact information about an accident is missing. The Leeway model accounts for the uncertainties in the position of an accident \cite{Breivik2008} by assigning initial positions from a normal distribution centered around an expected location of an accident. Particularly, the drifters are released continuously in a spatial area defined by initial distance $r_0=0.5 \sigma$ around the last known position (LKP), where $\sigma_r$ is the variance of drifters initial positions. The LKP refers to the possible location of an accident and is provided in terms of longitude and latitude for every drifter. In the experiments studied here drifters were released simultaneously. In general uncertainty in time could be included \cite{Breivik2008}.

To include the sub--grid variabilities of the wind and current velocities the Gaussian perturbations were added at every integration step $t_n$ and for the $k$th drifter: 
\begin{eqnarray}
v_{x,y}^{curr}(t_n,k)&=&v_{x,y}^{curr}+\xi_{x,y}^{curr},\\
w_{x,y}(t_n,k)&=&w_{x,y}+\xi_{x,y}^{wind},\label{Gaus_perturb}
\end{eqnarray}
where $\xi_{x,y}^{curr}$ and  $\xi_{x,y}^{wind}$ are the uncertainties chosen from a normal distributions $N(0,\sigma_{curr})$ and $N(0,\sigma_{wind})$ correspondingly. The standard deviation (std) of wind $\sigma_{wind}$ current $\sigma_{curr}$ were estimated using averages over the corresponding geographical domain. 

\section{Methods}\label{sec.3}

\subsection{Interpolation schemes}

The Eulerian velocity of drifter $\mathbf{V_E}$ was obtained using bilinear interpolation of the model zonal and meridional components at the position of a drifter. Also weighted linear time interpolation was used to interpolate current and wind velocities between two subsequent time stamps. In our simulations the integration step $\delta t=360 s$ was taken.
For the advection scheme the Runge Kutta second--order method was used with the Euler first--order trial step \cite{Tikhonov2011,Teukolsky1992}. 

\subsection{Clustering procedure} 
In the implementation of the algorithm the final positions of drifters were used for evaluation of a spatial separation between pairs of drifters. The separation distance was estimated from the haversine formula:
\begin{equation}
d=2 R_{\oplus}\arcsin{\sqrt{f_{\phi_i,\phi_j}+\cos\phi_i \cos\phi_j f_{\lambda_i,\lambda_j}}},\label{haversine}
\end{equation}
where $R_{\oplus}$ is the Earth radius, $i,j=1\ldots N_{dr}$ are drifter indices, $\lambda_i, \phi_i$ are longitudes and latitudes of the $i$th drifter and $f_{x_1,x_2}=\sin^2{\frac{x_1-x_2}{2}}$. The mean separation distance $D_{mean}$ was evaluated from Eq.~\ref{haversine} across all distinct pairs of drifters. 
Also an upper bound $D_{max}>D_{mean}$ was assigned to exclude mean separation distance that exceeds the size of a given geographical domain. Indeed, in the case $D_{mean}\ge D_{max}$ no clusters were searched. 

For the $i$th drifter a new cluster was formed by the inclusion of the $k$th neighbor provided that the condition $D_{ik}<D_{mean}$ was satisfied. The procedure was finalized after sorting all possible pairs of drifters. 

Application of the algorithm produced a set of spatially separated clusters. We classified a probability of containment ($POC$) of drifters per cluster as follows: $ POC=N_{k} /N_{dr}\times 100 \%$, where $N_k$ is a number of drifters inside the $k$th cluster and $N_{dr}$ is an ensemble size. The mean cluster trajectory (center of mass of the cluster) is defined as the time sequence of mean longitudes and latitudes across all cluster members calculated at every time step.

\section{Oceanographic and atmospheric data}\label{sec.4}
Six hourly $10$ m wind fields with $0.5^\circ \times 0.5^\circ $ spatial resolution were retrieved from the ECMWF. 

For typical SAR objects of small size \cite{Breivik2008} only top surface levels of ocean currents were used. In our study, we used oceanographic data from two operational models: the MFS with a horizontal resolution of about $1/16^\circ \times 1/16^\circ$ \cite{Pinardi2003,Tonani2008} and the AFS (horizontal resolution from $1/22^\circ$ to $1/45^\circ$) \cite{Oddo2006,Guarnieri2010}. In the vicinity of shorelines a sea--over--land interpolation procedure was implemented to provide a horizontal interpolation of the ocean currents towards coastal zone. 

\section{Case study I: experiments in the Adriatic sea}\label{sec6}
The stochastic simulations of an ensemble of $N_{dr}=\mathcal{O}(2000)$ drifters were performed using various simulation scenarios (see the summary of experiments in Table~\ref{tab1}). All drifters were released simultaneously from the position $42^{\circ} 35'$ N, $16^{\circ} $ E.  

The mean surface wind evaluated for a study period (Fig.~\ref{fig1}) strongly contributed to the north--eastern displacement of surface drifters in the area around the release position. Also the western Adriatic boundary current (Figs.~\ref{fig2}) was influencing the transport of near coast drifters southwards during the given period. 

In the first experiment (Fig.~\ref{fig3}) a significant part of an ensemble of drifters was carried by the western coastal flow initially southwards and later towards the inner southern Adriatic circulation. The majority of drifters were localized in the southern Adriatic (clusters $2$ and $4$), a small part of the ensemble (cluster $1$) was stranded on the western coast and the remaining drifters due to low velocities were localized inside the release zone (cluster $3$).  

The second experiment (Fig.~\ref{fig4}) was done for person in water (PIW) category (see Table~\ref{tab2}) with the atmospheric wind. During $21$ days of simulations the wind and currents influenced the dispersion of drifters across the whole Adriatic basin towards the eastern Adriatic coast: small size clusters (clusters $2,4$ and $5$) and isolated drifters were localized in the central Adriatic. As in the experiment without the wind (Fig.~\ref{fig3}) a fraction of stranded drifters was found inside the release area (cluster $1$). No large size clusters were formed in this case.   
\begin{figurehere}
\begin{center}
  \scalebox{0.3}{\includegraphics{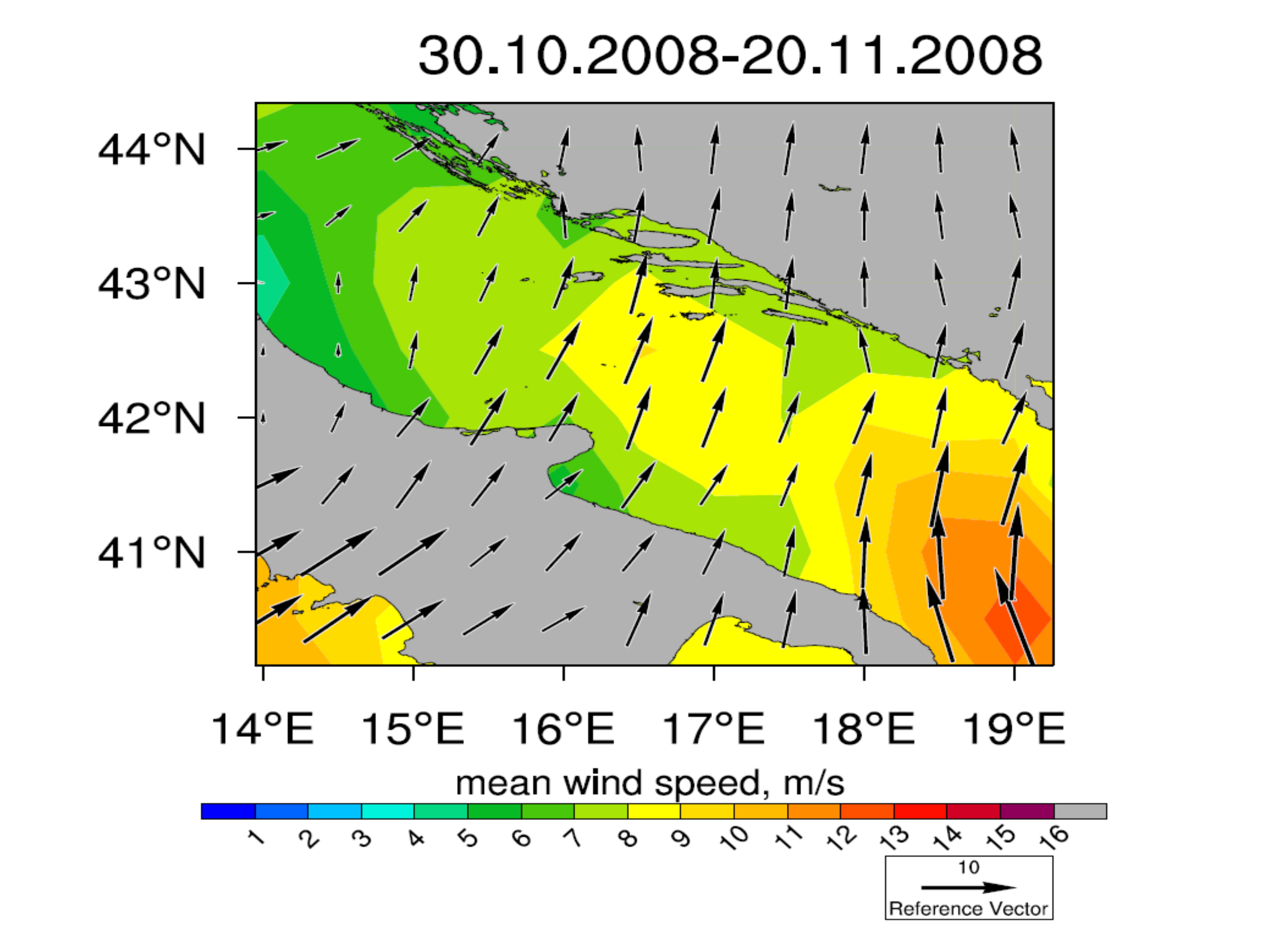}}
  \end{center}
  \caption{(Color online) (\textbf{a}) Magnitude, direction of mean [m/s] of the ECMWF winds for $30.10.2008-20.11.2008$.}\label{fig1}
\end{figurehere}
\begin{figurehere}
\begin{center}
  \scalebox{0.3}{\includegraphics{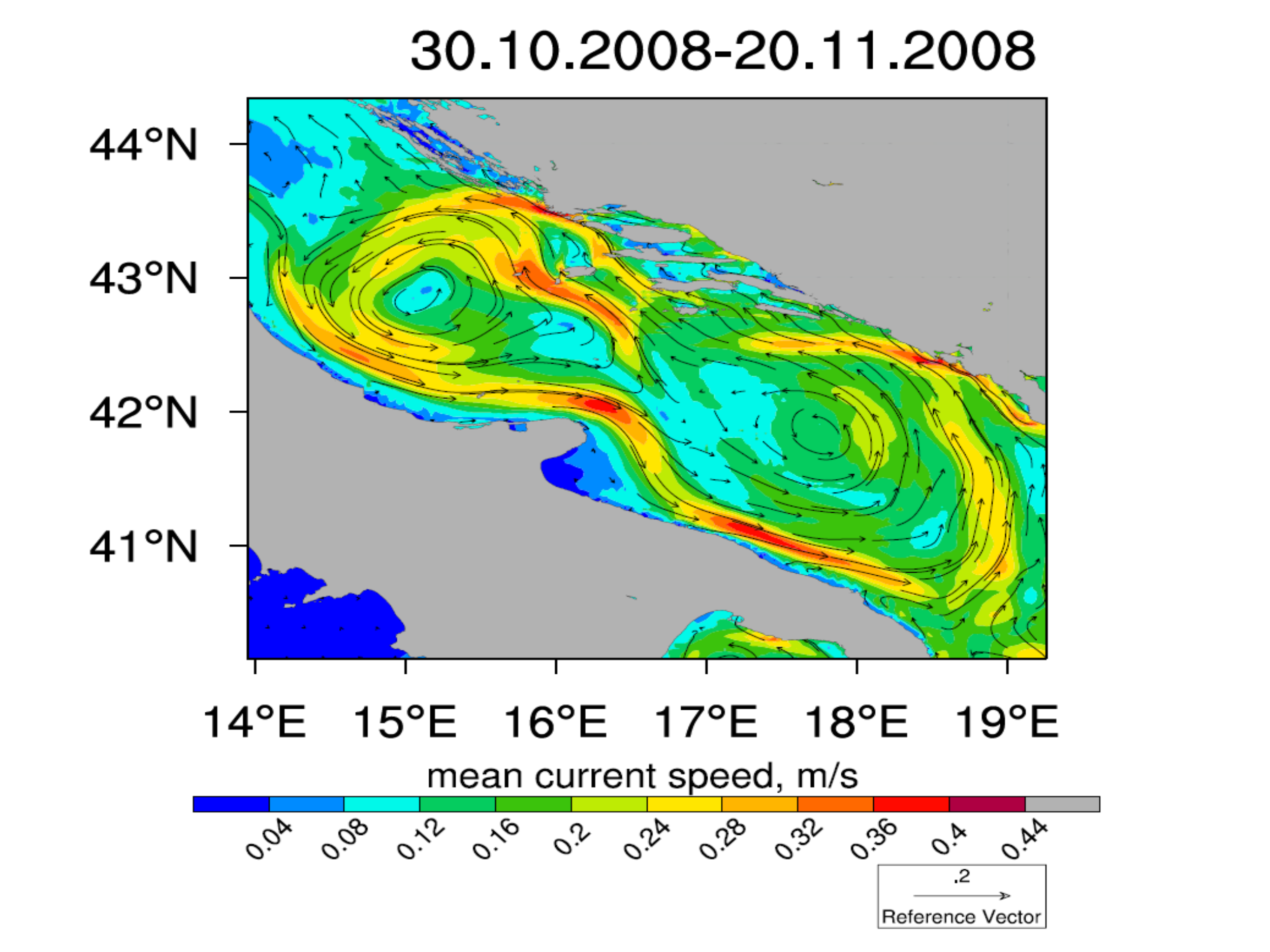}}
  \end{center}
\caption{(Color online) Magnitude and direction of the mean AFS currents for $30.10.2008-20.11.2008$: (\textbf{a}) mean surface currents [m/s].}\label{fig2}
\end{figurehere}
\end{multicols} 
 \begin{table}
\caption{\label{tab1} Summary of parameters for the simulation cases: experiment number, category of drifter ((-) sign  is indicated when no wind is used), initial radius $r_0$ of release around LKP [km] and depth [m].}
\vspace{0.2cm}
\begin{tabular}{l l l l}
  \hline
Exp. & Category & $r_0$ & Depth\\
 \hline
 $$ & $$ & $$ & $$ \\
 $1$ & - & $10$ & $0$\\
 $2$ & PIW (mean) & $10$ & $0$ \\
 $3$ & - & $10$ & $10$\\
 $$ & $$ & $$ & $$\\
\hline
\end{tabular}
\end{table}
\begin{multicols}{2}
In the third experiment (Fig.~\ref{fig5}) a current at the $10$--m depth level was used to account for possible scenarios of motion of submerged drifters. A higher clustering tendency was found in this case: the drifters were densely positioned inside three large clusters. The largest group of drifters (cluster $1$ with $POC>50\%$) was localized inside the southern Adriatic circulation. Also the western Adriatic current contributed towards the transport of a group of drifters (cluster $3$) to the southern end of the Salento peninsula. No stranded drifters were found apart from a set of drifters that did not leave the release area.  
%
\begin{figurehere}
  \scalebox{0.25}{\includegraphics{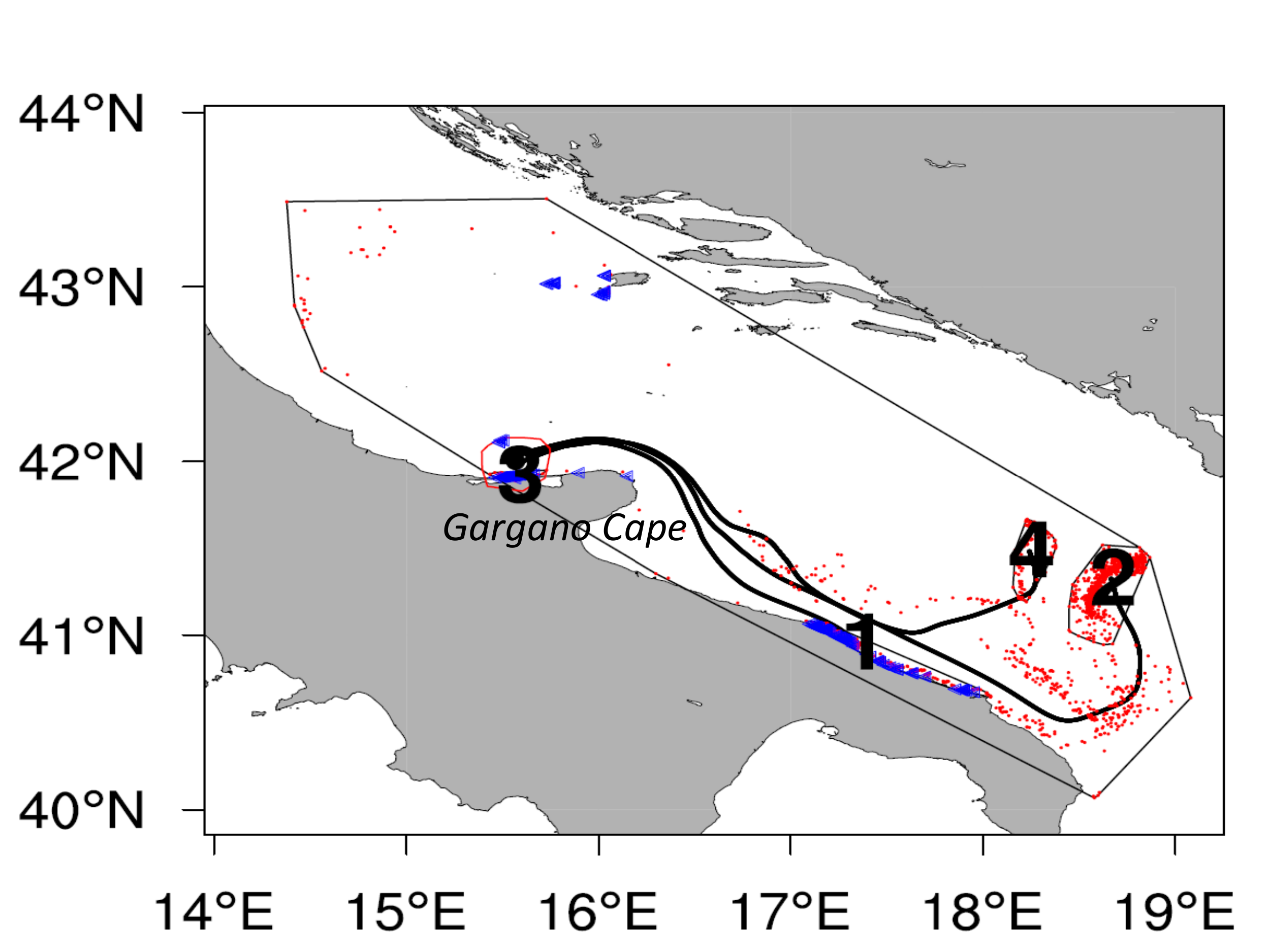}}
    \caption{(Color online) Leeway simulations using the AFS daily surface currents without wind: the LKP (big dot), final position of drifters (red dots), mean trajectories of drift for every cluster (black curve), cluster numbers and search area (black convex polygon) for every cluster, total area of search with all drifters inside (black convex polygon), drifters--outliers in open sea (red dots) and stranded drifters (blue triangles). Four clusters were found: ($1$) $POC=8.55\%$, ($2$) $POC=41.45\%$, ($3$) $POC=5.65\%$ and ($4$) $POC=22.85\%$.}\label{fig3}
\end{figurehere}
\begin{figurehere}
  \scalebox{0.25}{\includegraphics{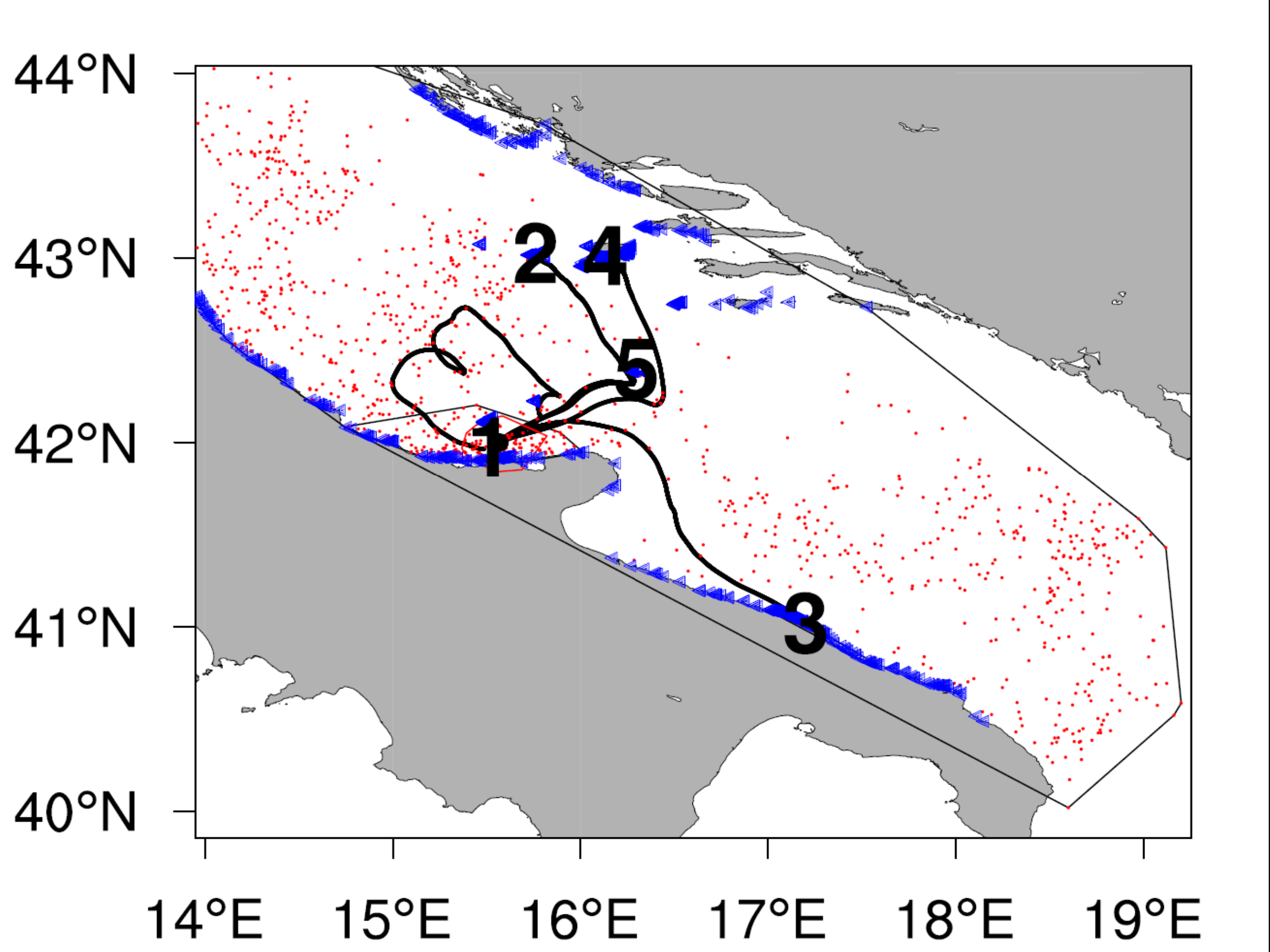}}
  \caption{(Color online) Leeway simulations with inclusion of $10$~m wind. Values for PIW (mean values) are used to parametrize the Leeway drift. Five clusters were formed: ($1$) $POC=17.1\%$, ($2$) $POC=6.9\%$, ($3$) $POC=5.9\%$, ($4$) $POC=6.6\%$ and ($5$) $POC=3.6\%$.}\label{fig4}
\end{figurehere}
\begin{figurehere}
  \scalebox{0.25}{\includegraphics{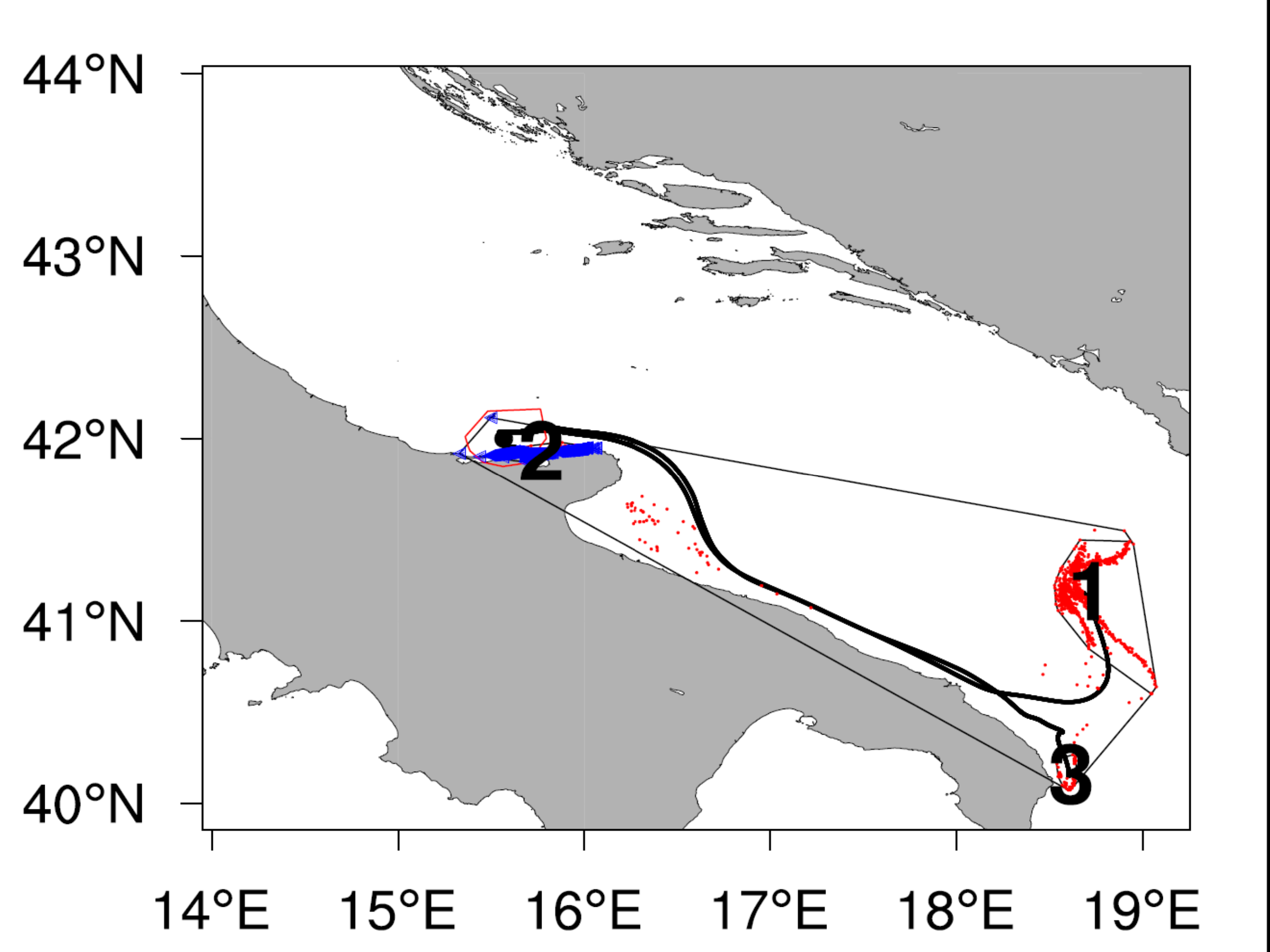}}
  \caption{(Color online) Leeway simulations for subsurface drift. The current at $\sim10$~m depth of the MFS model was used. Three clusters were formed: ($1$) $POC=50\%$, ($2$) $POC=42.6\%$ and ($3$) $POC=4\%$.}\label{fig5}
\end{figurehere}
\vspace{.5cm}
\section{Case study II: reconstruction of the Elba accident}\label{sec.5}
To set up initial conditions and define parameterization of an object drift the documented data about the accident were used \cite{Shcheki2013}: a person on an inflatable raft was lost in vicinity of coast of the Elba Island at $42^{\circ} 43' 60''$ N, $10^{\circ} 9' 5''$ E on $21.06.2009$ at $1:30$~UTC and after $34$ hours he was found alive at the position $42^{\circ} 22' 90''$ N, $9^{\circ} 53' 30''$ E in the open sea, it is known that at the time of rescue ($22.06.2009$ at $11:30$~UTC) the boat was partially deflated. 
\begin{figurehere}
\begin{center}
  \scalebox{0.25}{\includegraphics{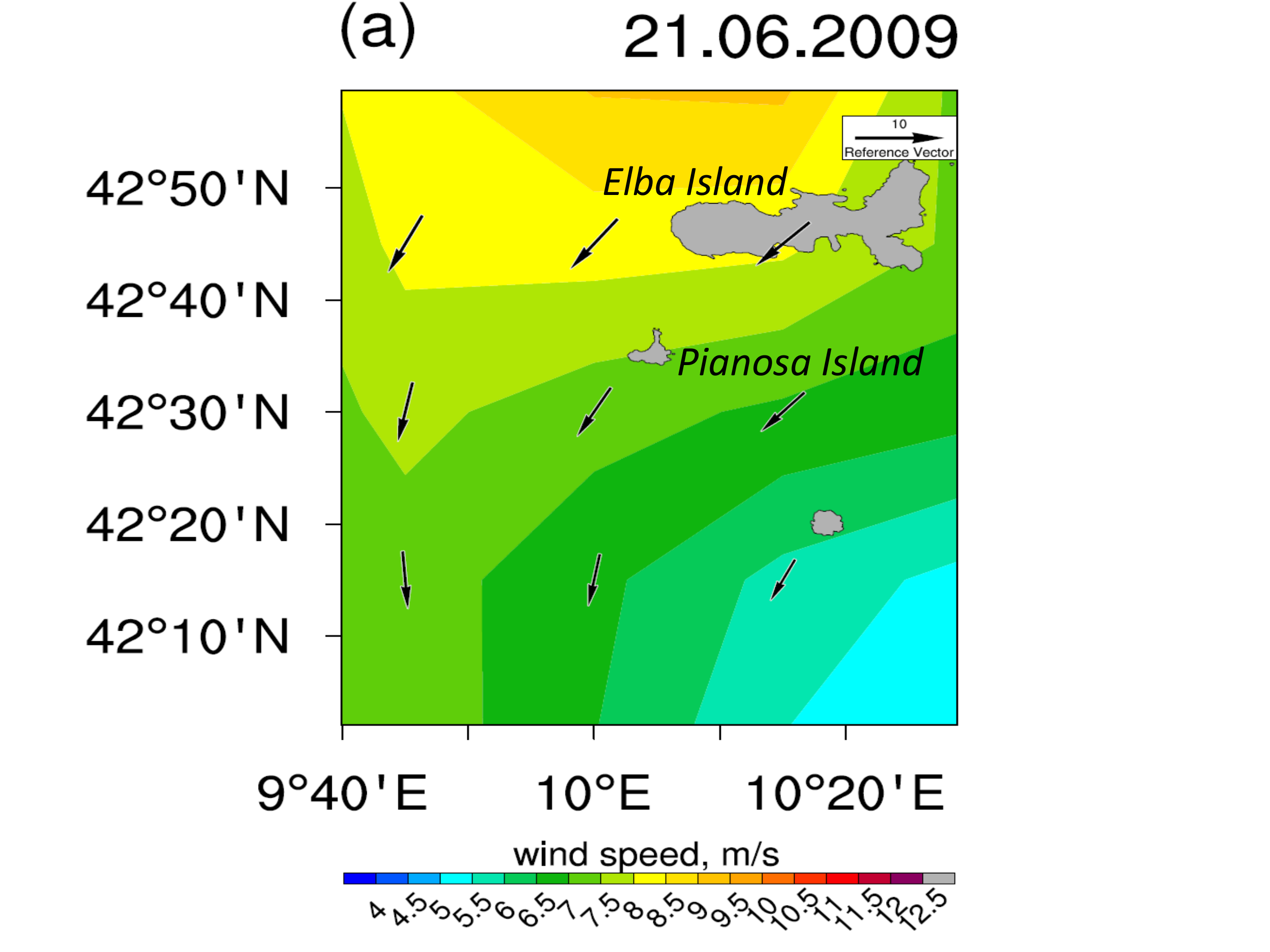}}
    \scalebox{0.25}{\includegraphics{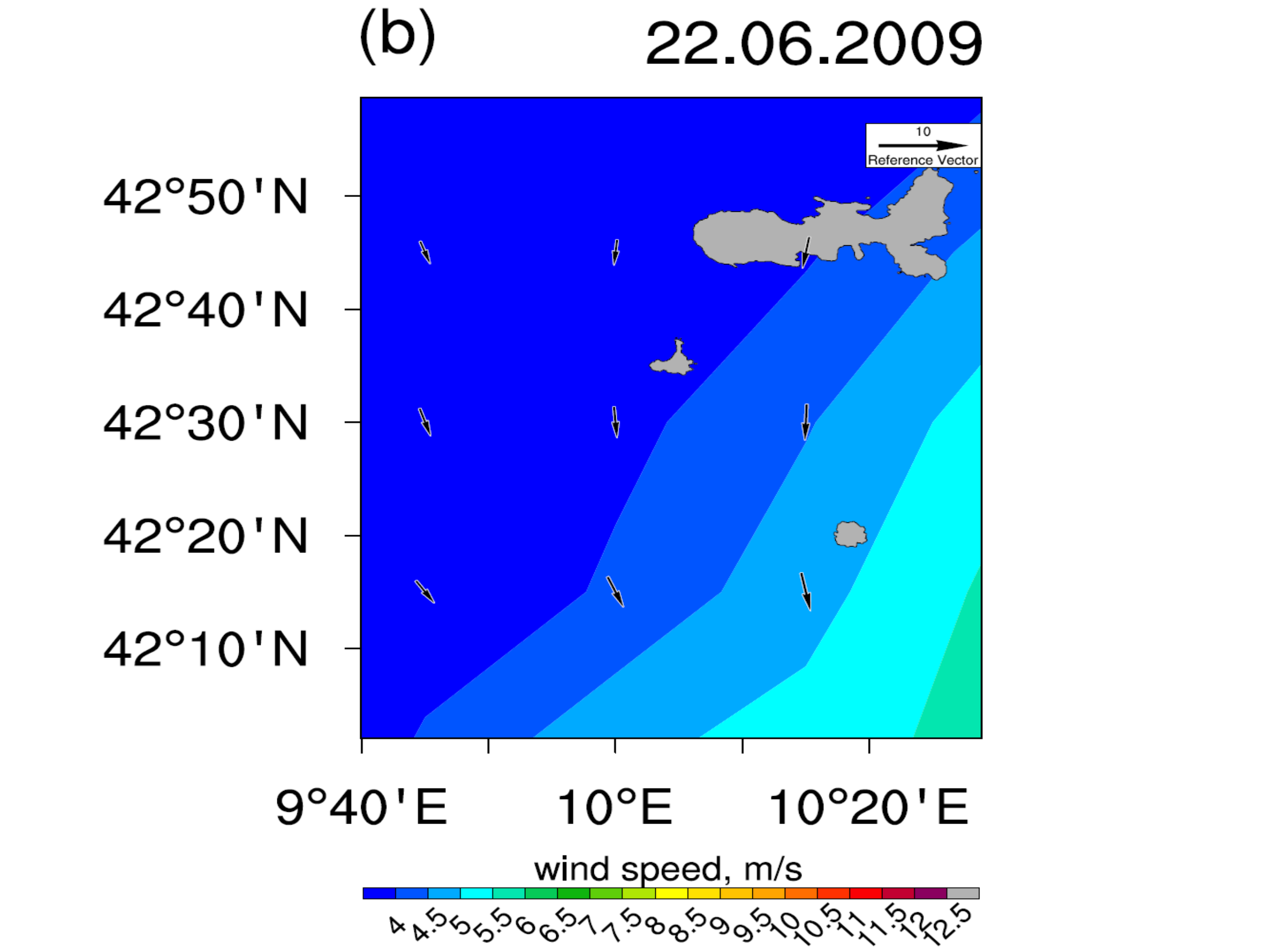}}
  \end{center}
  \caption{(Color online) daily wind direction and magnitude from the ECMWF for days: (\textbf{a}) $21.06.2009$ and (\textbf{b}) $22.06.2009$. }\label{fig6}
\end{figurehere}
\begin{figurehere}
\begin{center}
  \scalebox{0.25}{\includegraphics{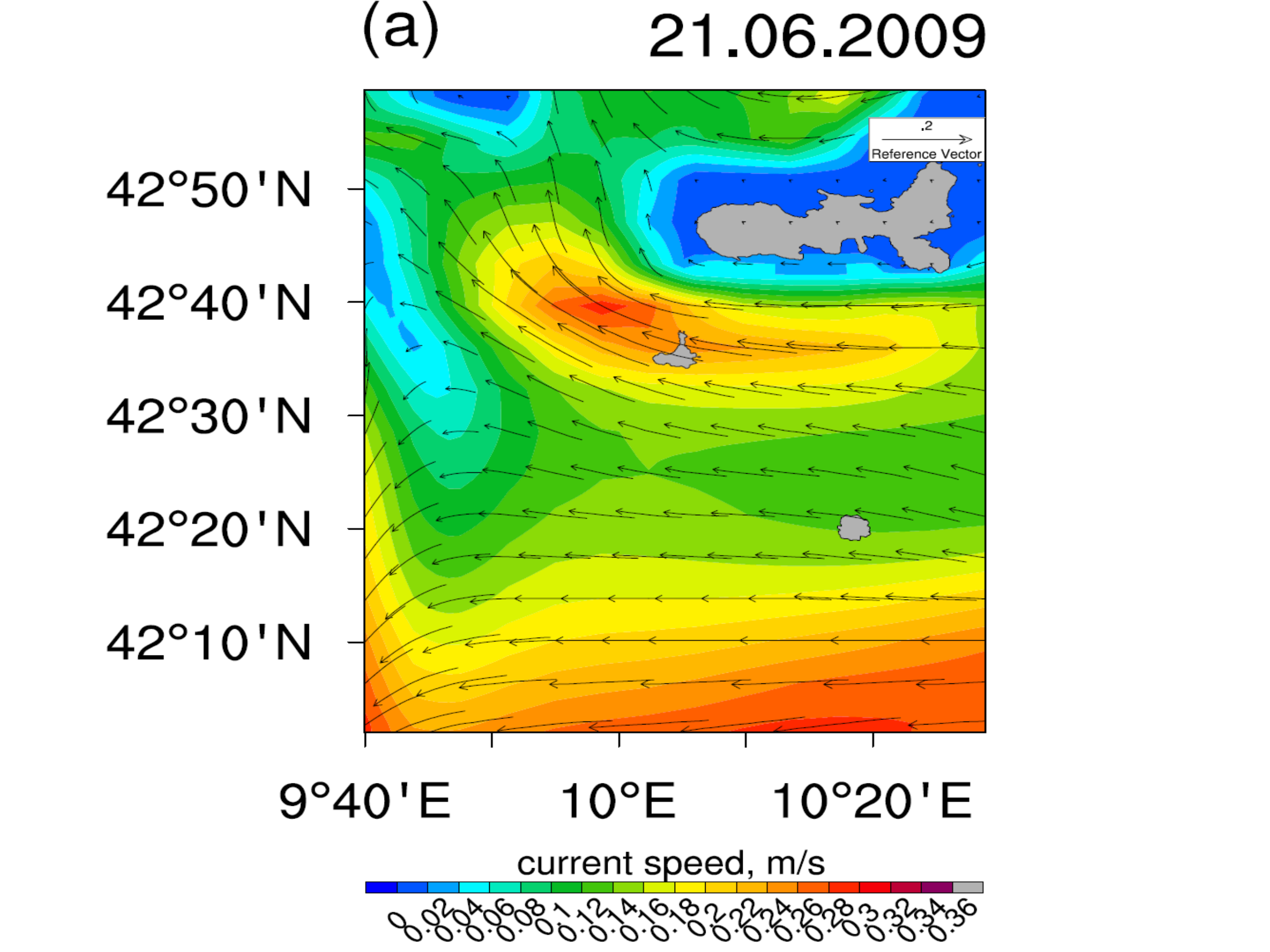}}
    \scalebox{0.25}{\includegraphics{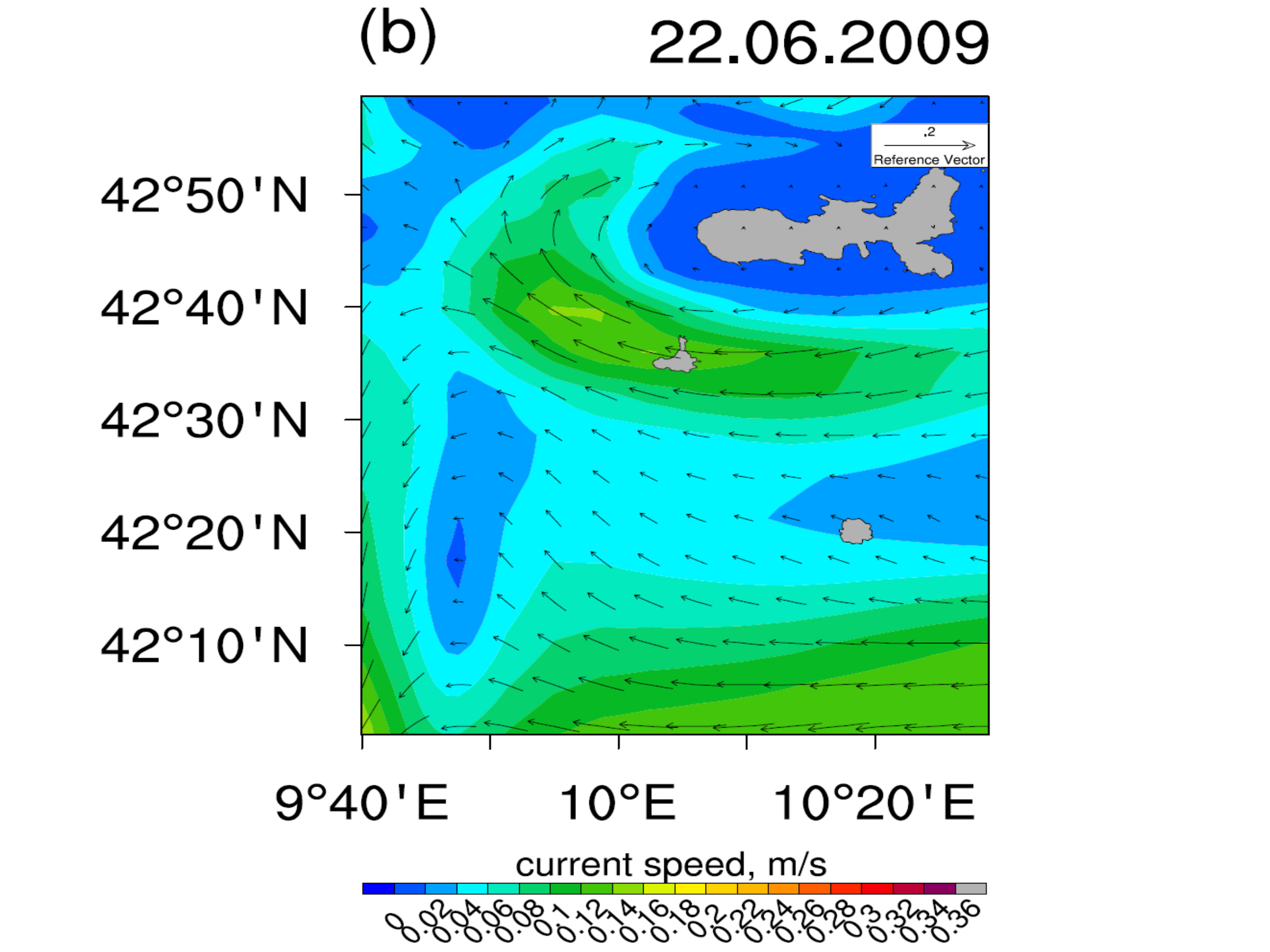}}
  \end{center}
  \caption{(Color online) daily surface currents direction and magnitude from the MFS for days: (\textbf{a}) $21.06.2009$ and (\textbf{b}) $22.06.2009$. }\label{fig7}
\end{figurehere}

Since the exact object parameterization was not known the simulations of drifters were performed using the categories similar to a PIW on an inflatable boat (see Table~\ref{tab2}). The ensemble of $N_{dr}\sim \mathcal{O}(3000)$ members was released simultaneously from the same LKP ($42^{\circ} 43' 60''$ N, $10^{\circ} 9' 5''$ E). The start date and time $30.10.2008,10:30 ~UTC$ of the release of drifters was the same in all the experiments. Drifters were released in a circle with radius $r_0=15$~km around LKP. The minimum number of drifters per cluster is $N_{min}=20$.
  
\begin{figure}[htp]
\centering
\begin{tabular}{cc}
   \includegraphics[width=60mm]{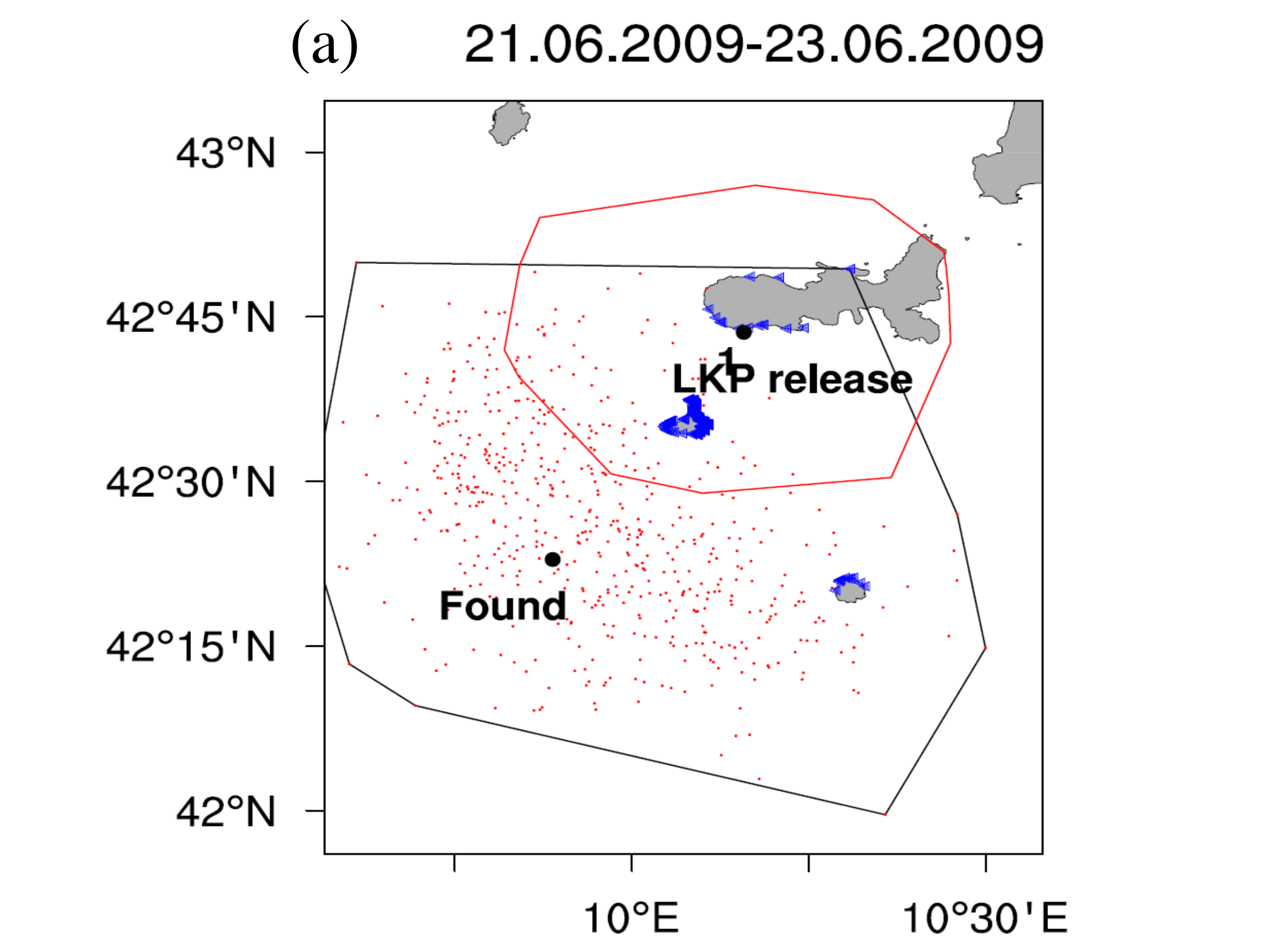}&
   \includegraphics[width=60mm]{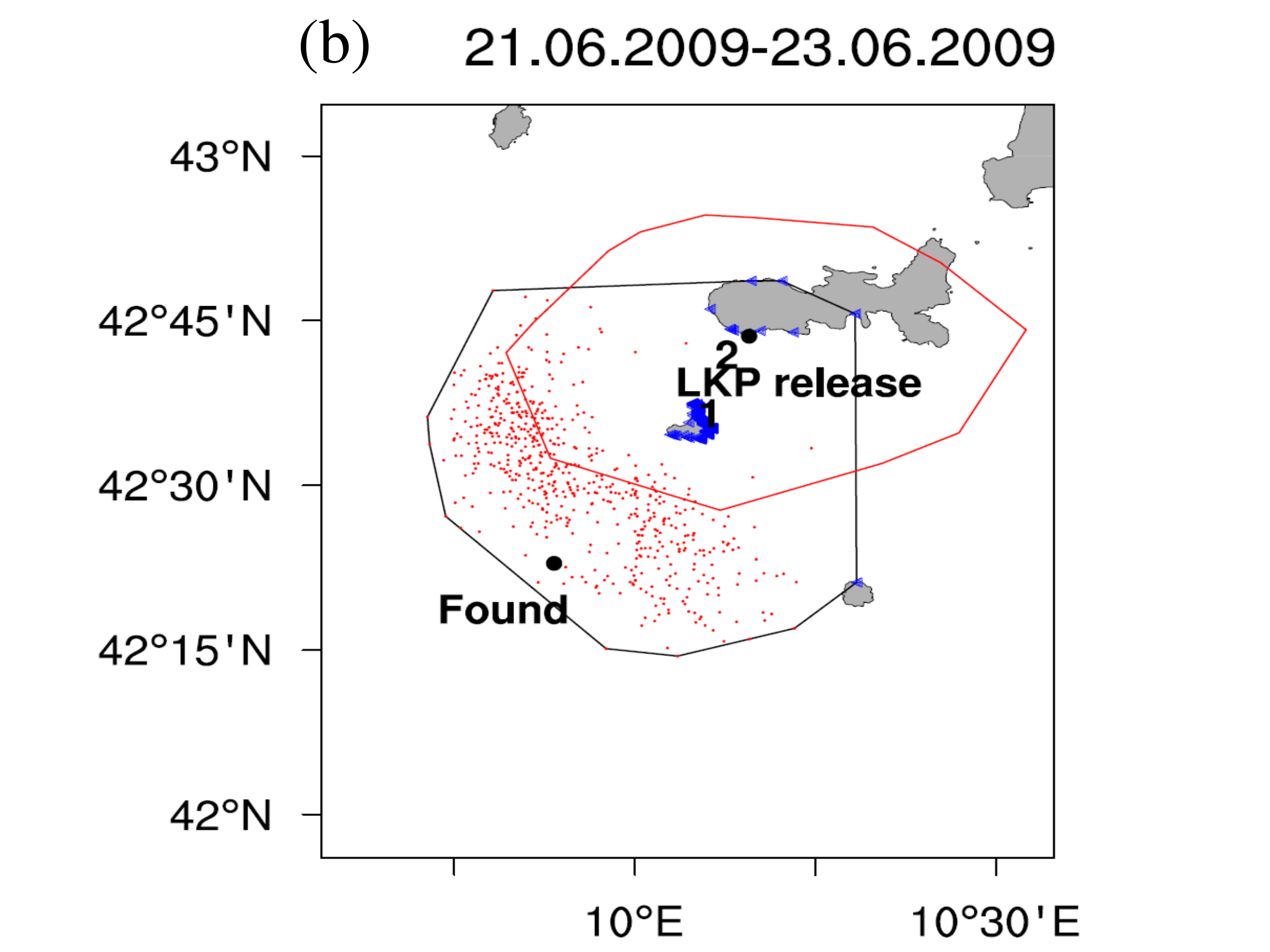}\\
   \includegraphics[width=60mm]{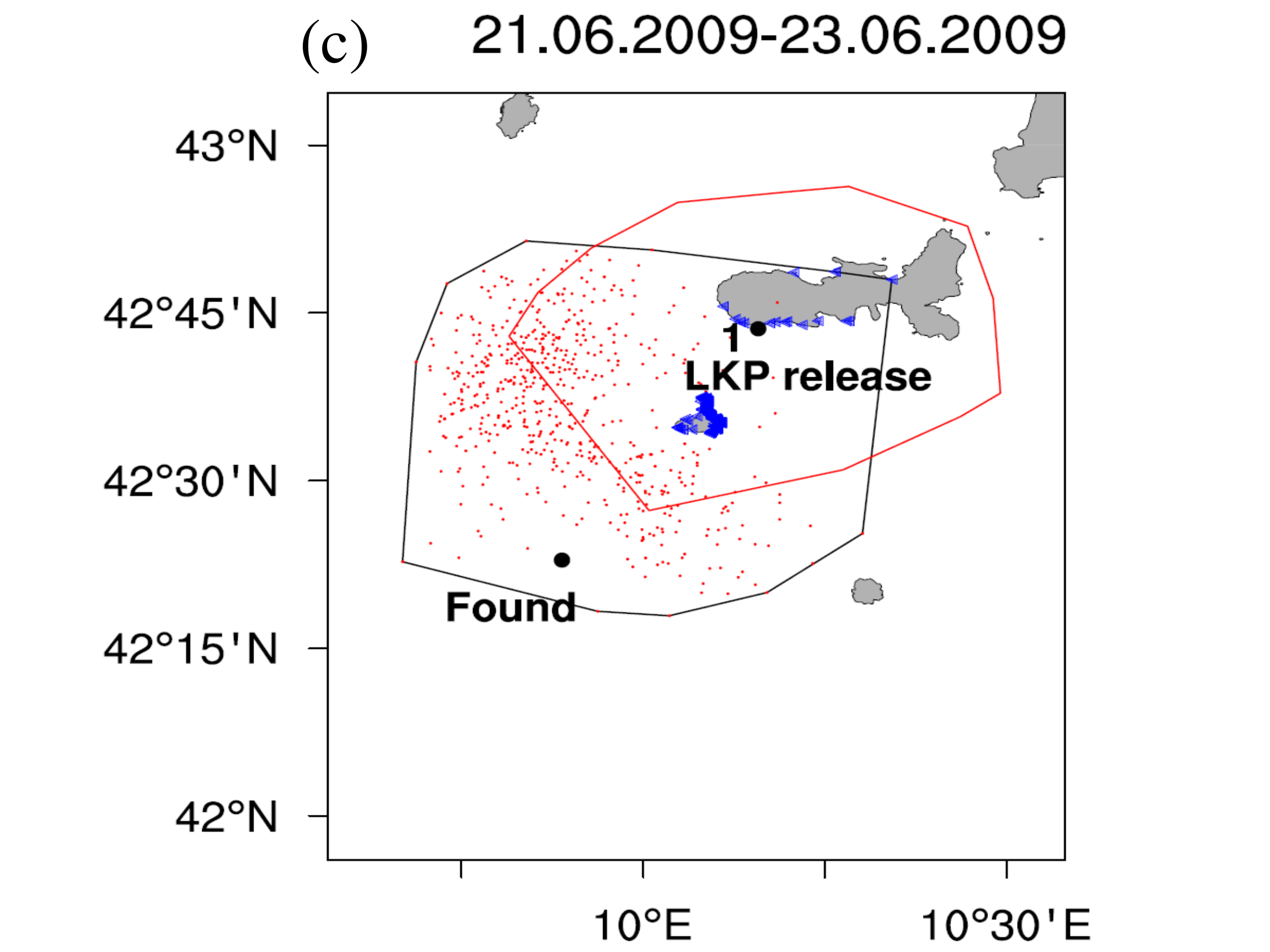}&
   \includegraphics[width=60mm]{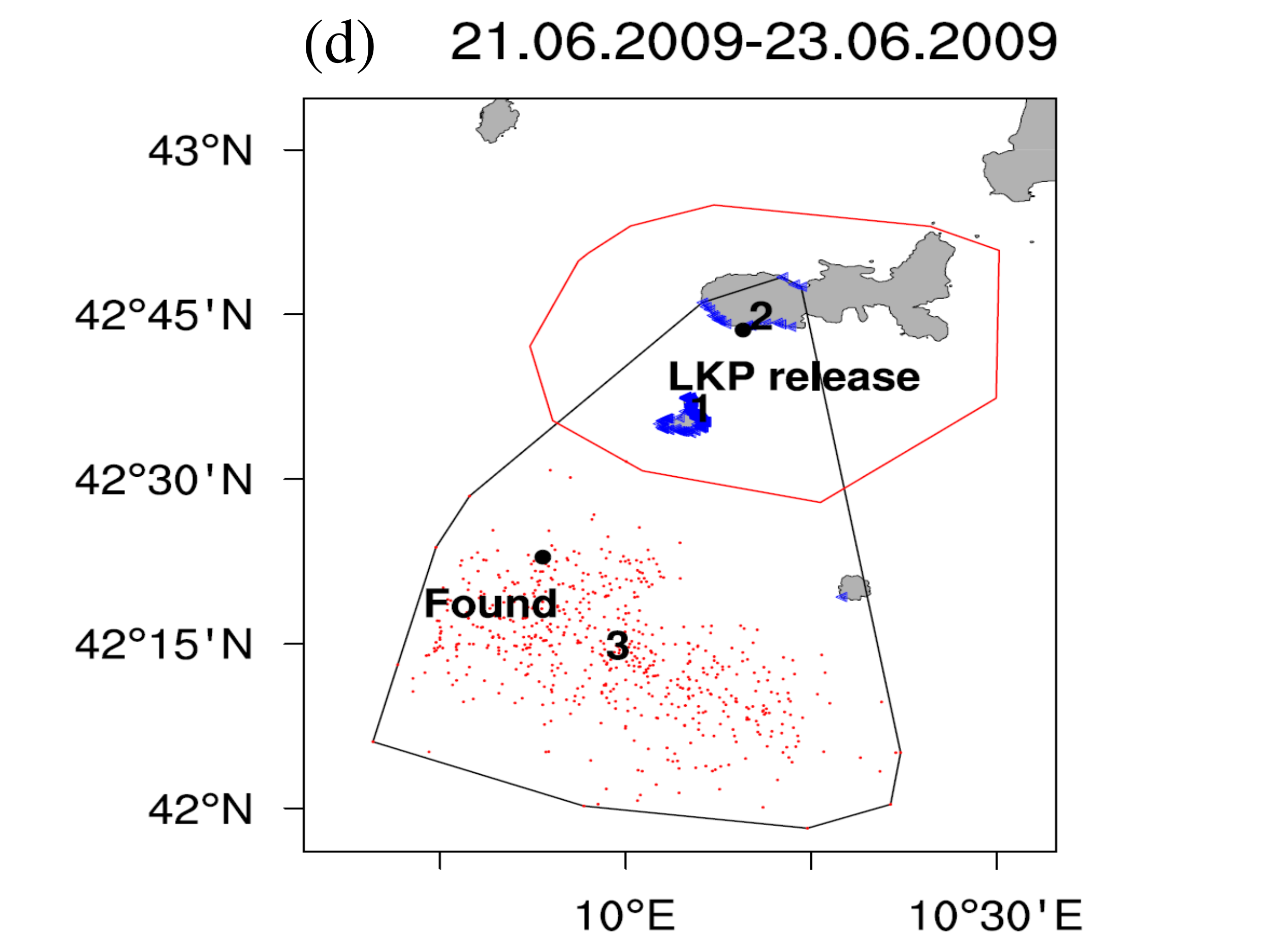}\\
   \includegraphics[width=60mm]{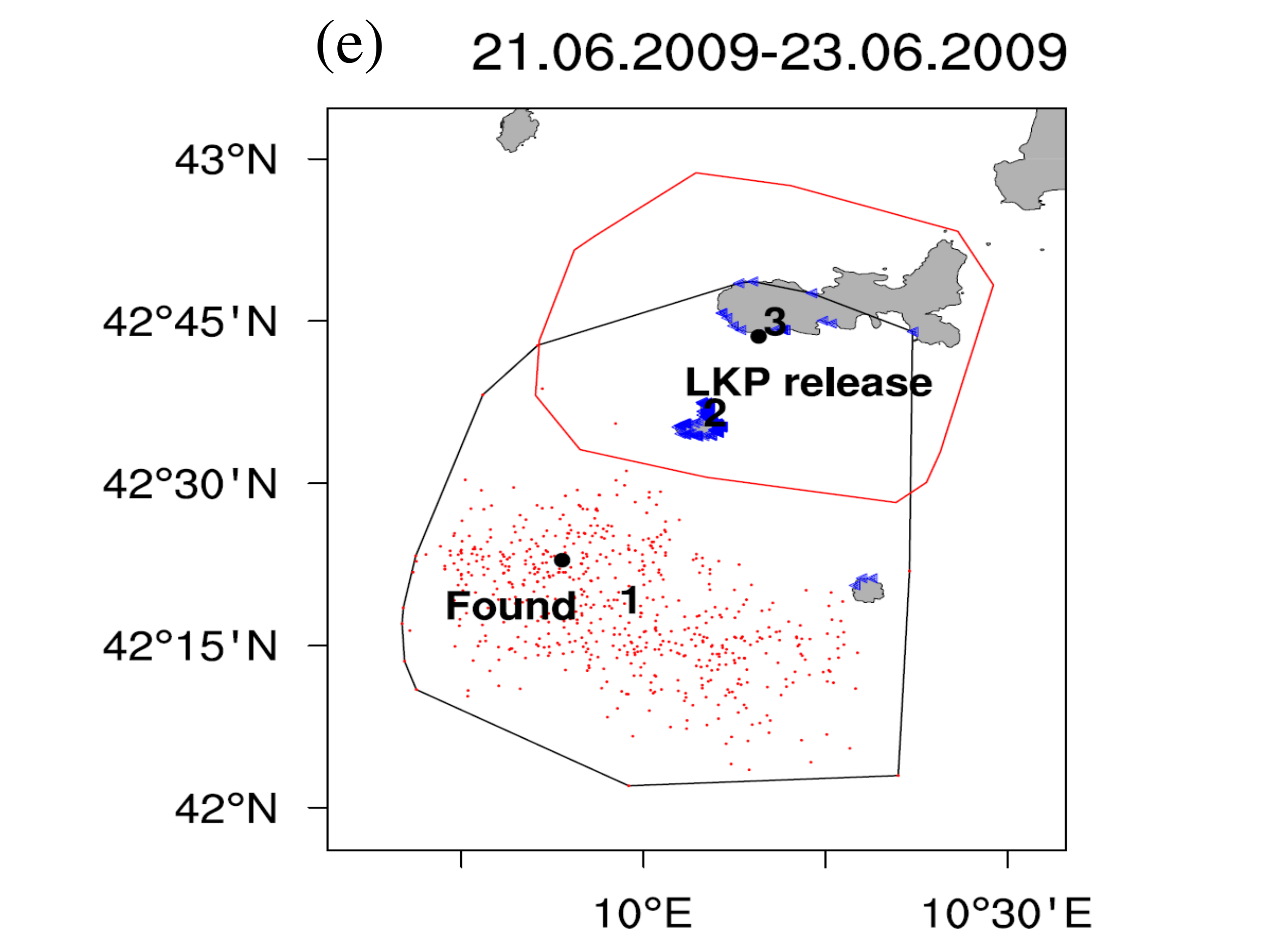}\\
  \end{tabular}
  \caption{(Color online) Leeway simulations for the Elba accident: LKP ($42^{\circ} 43' 60''$ N, $10^{\circ} 9' 50''$ E), position where the person was found  ($42^{\circ} 22' 90''$ N,  $9^{\circ} 53' 30''$ E ), final positions (red dots) in open sea and (blue triangles) stranded on coast, area of drifters release (red convex polygon), final area of search (black convex polygon). (\textbf{a}) Results for the a life raft without ballast. Clustering algorithm distinguishes a single cluster with all drifters included.
    (\textbf{b}) Results for a life raft with a deep ballast. Two clusters are found: ($1$) $POC=7.62\%$ and ($2$) $POC=92.1\%$. Cluster $1$ consists of stranded drifters. 
    (\textbf{c}) Results for a surfboard with person. Drifters form a single cluster.
    (\textbf{d}) Results of simulation for a sports boat. Three clusters are distinguished: ($1$) $POC=19.6\%$ (stranded drifters), ($2$) $POC=7.2\%$ (stranded drifters) and ($3$) $POC=72.8\%$ (open sea drifters).
    (\textbf{e}) Results for a sport fisher. Three clusters are found: ($1$) $POC=73.1\%$ (open sea drifters), ($2$) $POC=19.8\%$ (stranded drifters) and ($3$) $POC=6.6\%$ (drifters near to the LKP).
    }\label{fig8}
\end{figure}

Both the south--western wind on $21.06.2009$ (Fig.~\ref{fig6}~a) and the westward surface current (Fig.~\ref{fig7}~a) contributed to initial southwestern displacement of drifters with respect to the release area Figs.~\ref{fig8}. On the following day the decrease of the mean surface current (Fig.~\ref{fig7}~b) and change of the wind direction (Fig.~\ref{fig6}~b) favored a slight deflection of drift trajectories to the southeast (Figs.~\ref{fig8}).
In the first experiment (Fig.~\ref{fig8} a) the search area included the found position but the drifters were spread over large area and thus their average density around the found position was low. In the next two experiments (Fig.~\ref{fig8} b and c) an overall southwestern drift from the LKP was observed. Two clusters were found for the liferaft with a deep ballast with the largest ($POC>91\%$) formed by open sea drifters. In the last two experiments (Figs.~\ref{fig8} d and e) the southwards drift was more pronounced and trajectories more closely represented the observed real trajectory during the accident. The estimated search areas (Figs.~\ref{fig8} d and e) included the position of rescue with the largest clusters formed. 
\end{multicols}
\begin{multicols}{2}
To summarize the results of simulations: the trajectories of drifters with higher DWL and lower CWL (see Table~\ref{tab2}) more closely reproduced target trajectory in the accident. Generally, these drifters are draft--limited to an upper sea. Hence their trajectories were more influenced by the wind and wind--induced forcings during experiments. In contrast, the vessels with a deep draft were less affected by wind and drifted westwards from the Elba location. Due to details of ensemble initialization a significant fraction of drifters was found stranded on the neighboring island and a set of drifters with low initial velocities was stranded near to the LKP. In the experiments with draft limited drifters (Figs.~\ref{fig8} d and e) a larger separation between sets of coastal and open sea drifters was observed.   
\section{Conclusions}\label{sec8}
The stochastic simulations in the Adriatic sea identified most probable search areas with a high concentration of drifters: while a significant portion of ensemble members entered the central Adriatic pit another group of drifters moved with the western Adriatic coastal current and was either found on the coast or entered the southern Adriatic circulation. We observed high sensitivity of trajectories to the details of an initial ensemble such as category of drifters as well as choice of geophysical forcing. In all the experiments the central and southern Adriatic circulations were important contributions for the transport of drifters. These circulations mediated the detachement of drifters from the coast and further displacement inside the main Adriatic gyres. The clustering procedure helped finding very localized search areas with a high density of drifters in the experiment with subsurface drifters. Our results confirmed previous studies \cite{Veneziani2006} that highlighted an influence of underlying flow hyperbolicity on the transport of drifters near to the Gargano Cape. 

In the numerical reconstruction of the Elba accident the estimation of the most probable drift path was provided. The search area identified by clustering method matched the documented final position. The results showed that draft--limited drifters more closely reproduced observed final position during the accident.
\section{Appendix}
\subsection{Definitions of Leeway coefficients }
Here the explicit form of empirical relationship between Leeway components and magnitude of wind is provided. The DWL component $L_{d}$ and CWL component $L_{c}$ are linearly related to the wind magnitude $W_{10}=\sqrt{w_x^2+w_y^2}$ according to derived relationships \cite{AllenPlourde1999,Breivik2008}:
\begin{equation}
L_{d,c}=a_{d,c} W_{10} +b_{d,c}, \label{Leewaycoeff}
\end{equation}
where $a_d,b_d$ and $a_c,b_c$ are the linear regression coefficients for the DWL and CWL correspondingly. The regression coefficients $a_{d,~c},b_{d,~c}$ are obtained by adding noise terms to linear regression coefficient from experimental observations \cite{AllenPlourde1999}:
\begin{eqnarray}
a_{d,c}&=&\alpha_{d,c}+\zeta_{d,c}/20,\nonumber\\
b_{d,c}&=&\beta_{d,c}+\zeta_{d,c}/2,\label{leewayuncertainty}
\end{eqnarray}
where $\alpha_{d,c},\beta_{d,c}$ are regression slope and offset terms obtained from empirical data \cite{AllenPlourde1999}, The noise terms $\zeta_{d,c}$ are taken from a normal distribution with the experimental value of variance $\sigma_{d,c}$ obtained from the field data \cite{AllenPlourde1999}.
\subsection{Equations of motion of a drifter}
The equation of motion of a floating body in the sea is derived from the force balance equation \cite{Breivik2008}:
\begin{equation}
(m-m') \frac{d\mathbf{V_{dr}}}{dt}=\mathbf{F_{wave}}+\mathbf{F_{wind}}+\mathbf{F_{ocean}},\label{forcebalance}
\end{equation}
where $m$ and $m'$ are masses of drifter and an additive mass, $\mathbf{V_{dr}}$ is a drift velocity , $\mathbf{F_{wind}}, \mathbf{F_{wave}}$ and $\mathbf{F_{ocean}}$ are wind, wave and water drags correspondingly. 

Since observations show that ships and small size floaters \cite{Sorgard1998,Hodgins1998} reach terminal velocity rapidly an infinite acceleration and constant velocity during every time step is assumed. 
When the wave force is neglected \cite{Sorgard1998,Breivik2008} Eq.~\ref{forcebalance} with implicit expressions for wind $F_{wind}=C_{a} \rho_a A_a \|W_{10}\| \mathbf{W_{10}}$ and water drag $F_{ocean}= C_{o} \rho_o A_o ^2\|\mathbf{V_{L}}-\mathbf{dr/dt}\| (\mathbf{V_{L}}-\mathbf{dr/dt})$ takes the form \cite{Anderson1998}:
\begin{eqnarray}
0&=& C_a\rho_a A_a \|W_{10}\| \mathbf{W_{10}}\nonumber\\& &-C_{o}\rho_o A_o \left|\mathbf{V_{L}}-\mathbf{\frac{dr}{dt}}\right| \left[\mathbf{V_{L}}-\mathbf{\frac{dr}{dt}}\right],\label{massbalanceext}
\end{eqnarray} 
where $\mathbf{V_L}$ is the Lagrangian velocity of drifter, $\mathbf{\frac{dr}{dt}}$ is the drifter velocity, $C_{a,o}$ are the drag coefficients, $\rho_{a,o}$ are densities and $A_{a,o}$ are effective cross sectional areas of body exposed to air and water correspondingly. Thus the equation of motion of drifters can be written as follows \cite{Roehrs2012}:
\begin{equation}
\mathbf{\frac{dr}{dt}}=\mathbf{V_{L}} +\alpha\mathbf{W_{10}}, \label{eqmotion1}
\end{equation}
where the coefficient $\alpha=\sqrt{\rho_a A_a C_a/\rho_o A_o C_o}$ depends on the water/air constants. By expressing the Lagrangian velocity in terms of the Eulerian current and the Stockes drift $V_{St}$ we obtain the equation \cite{Roehrs2012}:
\begin{equation}
\mathbf{\frac{dr}{dt}}=\mathbf{V_{E}} +\beta\mathbf{W_{10}}, \label{eqmotion2}
\end{equation}
where $\beta=\alpha+\|V_{St}\|/\|\mathbf{W_{10}}\|$ is derived from the assumption that the wind direction and the Stockes drift are parallel \cite{Roehrs2012}. The Leeway model uses explicit substitution of the second term in (\ref{eqmotion1}) by the Leeway drift parametrized by wind (\ref{Leewaycoeff}).
\subsection{List of categories of drifters and Leeway coefficients}
The Leeway coefficients for the categories used in the experiments are provided here. The slope, offset and std for the CWL,~DWL are given in Table~\ref{tab2}.
\end{multicols}
 \begin{table}
\caption{\label{tab2} Leeway coefficients for the categories used in the simulations. The regression slope values  $\alpha_{d,c}$ $[\%]$, the offset $\beta_{d,c}$ [cm/s] and the standard deviations $\sigma_{d,c}$ [cm/s] are shown for the DWL and the left/right CWL components. Values are provided from \cite{AllenPlourde1999}.}
\vspace{0.2cm}
\begin{tabular}{l l l l l l l l l l }
  \hline
  $$ & $$ & $$ & $$  & $$ & $$ & $$ & $$ & $$ & $$\\
Object name & $ $& DWL & $ $ & right & CWL& $ $& left & CWL &$ $\\
$$ & $ $& $ $ & $ $ & $ $ & $ $& $ $&$ $& $ $ &$ $\\
$$ & $ \alpha_{d}$& $\beta_{d}, $ & $\sigma_d $ & $\alpha_{c} $ &$ \beta_{c},$& $\sigma_c $&$\alpha_{c} $& $\beta_{c}$ &$\sigma_c $\\
 \hline
  $$ & $$ & $$ & $$  & $$ & $$ & $$ & $$ & $$ & $$\\
  Life raft, no ballast &$3.7$ &$0$& $12$&$1.98$& $0$& $9.4$&$-1.98$& $0$ &$9.4$ \\
  Life raft, deep ballast &$3.52$ &$-2.5$& $6.1$&$0.62$& $-3$& $3.5$&$-0.45$& $-0.2$& $3.6$\\
  Surf board with person&$1.93$ &$0$ &$8.3$&$0.51$& $0$& $6.7$&$-0.51$& $0$& $6.7$\\
  Sport-boat & $6.54$& $0$ &$3.0$&$2.19$& $0$& $2.8$&$-2.19$& $0$& $2.8$\\
  Sport-fisher & $5.55$& $0$& $3.3$&$2.27$& $0$& $3$&$-2.27$& $0$& $3$\\
  PIW, (mean) &$0.96$& $0$ &$12.1$&$0.54$ &$0.0$& $9.4$&$-0.54$& $0.0$& $9.4$ \\
  $$ & $$ & $$ & $$  & $$ & $$ & $$ & $$ & $$ & $$\\
  \hline
  \end{tabular}
\end{table}
\begin{multicols}{2}
\section{Acknowledgements} \label{sec10}
The authors are grateful to N. Pinardi, G. Coppini and P. Oddo for scientific guidance and provision of the research facilities. The authors thanks S. Ciliberti, M. Mancini, E. Calcagnile, G. Verri, G. Mannarini, R. Lecci, S. Creti and other colleagues from Centro Mediterraneo per Gambiamenti Climatici for their kind support. We acknowledge support from the European Territorial Cooperation Programm \enquote{Ionian Integrated marine Observatory} and the project \enquote{Technology for Situational Sea Awareness} funded by National Operative Programm \enquote{Research and Competition}.

\end{multicols}
\end{document}